\documentclass[12pt,twocolumn]{emulateapj}

\tightenlines

\usepackage{ulem} 
\usepackage{color} 
\usepackage{graphicx}
\usepackage{epsfig,amsmath}
\usepackage{times}

\makeatletter

\newcommand{\Rmnum}[1]{\expandafter\@slowromancap\romannumeral #1@}
\makeatother

\newcount\listnorom
\newcommand\listromanDE{\global\advance \listnorom by 1
{\lowercase\expandafter{(\romannumeral\listnorom)}\ }}

\def\lsim{\raise0.3ex
  \hbox{$<$\kern-0.75em\raise-1.1ex\hbox{$\sim$}}\,}
\def\gsim{\raise0.3ex
  \hbox{$>$\kern-0.75em\raise-1.1ex\hbox{$\sim$}}\,}

\newcommand{\pairs}{e$^+$--e$^-$}
\newcommand{\fHe}{f_\mathrm{He}}
\newcommand{\fSN}{f_\mathrm{SN}}

\newcommand{\crhydro}{{\it CR-hydro-NEI}}

\newcommand{\HESS}{{\it HESS}}
\newcommand{\CC}{core-collapse}
\newcommand{\Inj}{\chi_\mathrm{inj}}

\newcommand{\SunMyr}{$\Msun$\,yr$^{-1}$}

\newcommand{\RS}{reverse shock}

\newcommand{\pmax}{p_\mathrm{max}}

\newcommand{\dMdt}{\mathrm{d}M/\mathrm{d}t}
\newcommand{\Vwind}{V_\mathrm{wind}}

\newcommand{\SigWind}{\sigma_\mathrm{wind}}
\newcommand{\cutoff}{\alpha_\mathrm{cut}}

\newcommand\Msun{\mathrm{M}_{\odot}}

\newcommand{\Kep}{K_\mathrm{ep}}

\newcommand{\tSNR}{t_\mathrm{SNR}}
\newcommand{\dSNR}{d_\mathrm{SNR}}
\newcommand{\EnSN}{E_\mathrm{SN}}
\newcommand{\Mej}{M_\mathrm{ej}}

\newcommand{\SA}{semi-analytic}

\newcommand{\NEI}{non-equilibrium ionization}

\newcommand{\CD}{contact discontinuity}
\newcommand{\xx}[1]{\!\times\!10^{#1}}

\newcommand{\kmps}{km s$^{-1}$}

\newcommand{\gamray}{$\gamma$-ray}

\newcommand{\Valf}{v_A}
\newcommand{\alf}{Alfv\'en}

\newcommand{\falf}{f_\mathrm{alf}}
\newcommand{\SNRJ}{SNR RX J1713.7-3946}
\newcommand{\SNRJm}{RX J1713.7-3946}

\newcommand{\SC}{self-consistent}
\newcommand{\SCly}{self-consistently}

\newcommand{\be}{\begin{eqnarray}}
\newcommand{\ee}{\end{eqnarray}}

\newcommand{\rel}{relativistic}

\newcommand{\syn}{synchrotron}

\newcommand{\pdecay}{$\pi^0$-decay}
\newcommand{\IC}{inverse-Compton}
\newcommand{\brem}{bremsstrahlung} 
\newcommand{\brems}{bremsstrahlung} 

\newcommand{\mrm}{\mathrm}
\newcommand{\tna}{\,\tablenotemark{a}}
\newcommand{\tnb}{\,\tablenotemark{b}}
\newcommand{\tnc}{\,\tablenotemark{c}}
\newcommand{\tnd}{\,\tablenotemark{d}}
\newcommand{\tne}{\,\tablenotemark{e}}

\newcommand{\tnac}{\,\tablenotemark{$\ast$}}

\newcount\Inum
\newcount\IInum
\newcount\IIInum
\newcount\IVnum
\def\I{\global\multiply\IInum by 0 \global\multiply\IIInum by 0
            \global\multiply\IVnum by 0 \global\advance \Inum by 1
            {\the\Inum. }}
\def\II{\global\multiply\IIInum by 0\global\multiply\IVnum by 0
       \global\advance \IInum by 1 {\the\Inum.\the\IInum. }}
\def\III{\global\multiply\IVnum by 0\global\advance \IIInum by 1
            {\the\Inum.\the\IInum.\the\IIInum. }}
\def\IV{\global\advance \IVnum by 1
            {\the\IVnum. }}
\begin{document}

\title{A \textit{CR-hydro-NEI} Model of Multi-wavelength 
Emission from the Vela Jr. Supernova Remnant (SNR RX J0852.0-4622)}

\author{
Shiu-Hang Lee\altaffilmark{1},
Patrick O. Slane \altaffilmark{2},
Donald C. Ellison\altaffilmark{3},
Shigehiro Nagataki\altaffilmark{1}
and Daniel J. Patnaude \altaffilmark{2}
}

\altaffiltext{1}{Yukawa Institute for Theoretical Physics, Kyoto University, Oiwake-cho Kitashirakawa, Sakyo-ku, Kyoto 606-8502, Japan;
lee@yukawa.kyoto-u.ac.jp; nagataki@yukawa.kyoto-u.ac.jp}
\altaffiltext{2}{Harvard-Smithsonian Center for Astrophysics, 
60 Garden Street, Cambridge, MA 02138, U.S.A.;
slane@cfa.harvard.edu; dpatnaude@cfa.harvard.edu}
\altaffiltext{3}{Physics Department, North Carolina State
University, Box 8202, Raleigh, NC 27695, U.S.A.;
don\_ellison@ncsu.edu}

\slugcomment{Received 28 December 2012; Accepted 18 Feb 2013}

\begin{abstract}
Based largely on energy budget considerations and the observed cosmic-ray (CR) ionic composition, supernova remnant (SNR) blast waves are the most likely sources of CR ions with energies at least up to the ``knee" near $10^{15}$\,eV.
Shocks in young shell-type TeV-bright SNRs are surely producing TeV particles, but the emission could be dominated by ions producing \pdecay\ emission or electrons producing \IC\ (IC) gamma-rays. Unambiguously identifying the GeV-TeV emission process in a particular SNR will not only help pin down the origin of CRs, it will add significantly to our understanding of the diffusive shock acceleration (DSA) mechanism and improve our understanding of supernovae and the impact SNRs have on the circumstellar medium. In this study, we investigate the Vela Jr.~SNR, an example of TeV-bright non-thermal SNRs. We perform hydrodynamic simulations coupled with non-linear DSA and non-equilibrium ionization near the forward shock (FS) to confront currently available multi-wavelength data. We find, with an analysis similar to that used earlier for \SNRJ, that \SCly\ modeling the thermal X-ray line emission with the non-thermal continuum in our one-dimensional model strongly constrains the fitting parameters, and this leads convincingly to a leptonic origin for the GeV-TeV emission for Vela Jr. This conclusion is further supported by applying additional constraints from observation, including the radial brightness profiles of the SNR shell in TeV gamma-rays, and the spatial variation of the X-ray synchrotron spectral index. We will discuss implications of our models on future observations by the next-generation telescopes.
\end{abstract}
%
%


\section{Introduction}
Young shell-type supernova remnants (SNRs) with strong 
non-thermal emission, such as RX J1713.7-3946, HESS J1731-347 and Vela Jr., have been 
in the spotlight in recent years due to their presumed close relation to the origin of Galactic cosmic rays (CRs). These remnants show high luminosities in TeV gamma-rays and synchrotron dominated X-rays with no sign of thermal emission from the ejecta or shell. 
Vela Jr.\footnote{We note that the Vela Jr.~SNR is also referred to as G266.2-1.2.}~is 
a typical example of a TeV-bright SNR where emission in the X-ray band is strongly non-thermal.
It lies along the line-of-sight of the crowded Vela complex, comprising the Vela SNR, the Vela-X pulsar wind nebula (PWN), the Vela pulsar, the Pencil Nebula, and the Puppis A SNR. 
While Vela Jr.~was first discovered by the \textit{ROSAT} satellite at X-ray energies above $\sim 1.3$\,keV
\citep{Aschenbach1998}, it is coincident with the southeastern part of the more diffuse Vela SNR and is totally obscured by the latter in the soft X-ray band. 
The age of and distance to this remnant have been debated with the current most-agreed-on values being $d_\mathrm{SNR} \gtrsim 750$~pc and 
$t_\mathrm{age} \sim 1700 - 4300$~yr 
\citep[e.g.,][]{Slane2001,Katsuda2008}. 
The progenitor has been suggested to be the core-collapse of a massive star due to the discovery of a central compact object 
AX J0851.9-4617.4 near the center of the SNR \citep{Slane2001} with a consistent age and distance \citep{Kargaltsev2002}, although no pulsation has yet been detected. 

Detection of the 1.157~MeV radioactive decay line of the short-lived $^{44}$Ti along the same line-of-sight has been claimed \citep{Iyudin1998}, suggestive of an extremely short distance of about 200~pc, and a young age of around 700~yr. However, the line detection suffers from low statistical significance at the $2-4 \sigma$ level \citep{Schonfelder2000}, and from the non-detection of the associated Ca(Sc) K$\alpha$ line in X-rays by the \textit{Advanced Satellite for Cosmology and Astrophysics} (\textit{ASCA}) GIS \citep{Slane2001} and \textit{Suzaku} XIS \citep{Hiraga2009} imaging spectrometers. Moreover, using the distance and age deduced directly from proper-motion measurements \citep{Katsuda2008}, it can be shown that the required initial mass of $^{44}$Ti has to be $\sim$ 1~$\Msun$, orders of magnitude larger than that expected by supernova nucleosynthesis models. 
Hence, we consider the association of the $^{44}$Ti flux with Vela Jr.~unlikely. 

Sharp synchrotron filamentary structures in X-rays have been resolved by \textit{Chandra} in the bright northwestern shell \citep{Bamba2005}. An explanation of the small effective width of these filaments by fast synchrotron cooling of CR electrons requires a local magnetic field strength of about 100~$\mu$G 
\citep[e.g.,][]{Berezhko2009}, probably indicating magnetic field amplification (MFA) induced by efficient CR ion acceleration 
\citep*[e.g.,][]{Bell2004,VBE2008,BOE2011}. 

The radio remnant \citep{Combi1999, Stupar2005} is 
relatively faint ($\sim 30-50$~Jy at 1~GHz) with a flat spectral index $\sim 0.3$. The morphology is ring-like and shows a good overall correlation with the X-ray remnant, suggesting a common synchrotron origin for the 
radio and X-ray emission.  Spatially extended emission in the GeV and TeV bands has also been discovered by \textit{Fermi LAT} \citep{Tanaka2011} and \textit{H.E.S.S.} 
\citep{Aharonian2007b} along the line-of-sight. The gamma-ray spectra can be described by a single power-law with index $\sim 1.85$ and a cutoff energy at a few TeV. As we will discuss in more detail below, Vela Jr.~shares some striking similarities with another well-studied non-thermal SNR \SNRJm\ in both spectral and morphological characteristics. We believe these similarities are important for understanding the physical nature of Vela Jr.

Recently, \citet{Tanaka2011} have fitted the broadband spectra of Vela Jr.~using both hadronic (\pdecay\ dominated) and leptonic 
(IC dominated) scenarios. They assumed single power-law spectra with exponential cut-offs for the underlying proton and electron distributions. To obtain the best fits, they varied the spectral indices of the particles and the environmental parameters including the downstream (DS) B-field strength and gas density. %
They concluded that reasonable spectral fits can be obtained under both scenarios by invoking different parameter choices.
In some cases, however, the interpretation of their parameters can be non-trivial, such as an unrealistically large energy in the accelerated proton population ($\sim 10^{51}$~erg) in the hadronic model for a low-density ($n \sim 0.01$~cm$^{-3}$) ambient medium,  and a small downstream B-field ($B_2 \sim 10~\mu$G) required in the leptonic case which seems to contradict with the sharpness of the X-ray filaments found in the brighter NW rim (but see our arguments for such a possibility in Section~\ref{conclude}).

While DSA models predict the acceleration of both electrons {\it and} ions, determining the total energy appearing in relativistic particles depends crucially on the correct interpretation of which particles are responsible for any observed gamma-ray emission. In addition, models for which the gamma-ray emission is produced predominantly by hadrons necessarily produce observable signatures in the luminosity and ionization state of the thermal X-ray emission. 
We believe that disentangling the emission origin requires a 
more \SC\ SNR model, as we have developed and used to model \SNRJ\ 
\citep[e.g.,][and references therein]{ESPB2012,LEN2012}. In our \crhydro\ model, the thermal X-ray emission is determined \SCly\ using the dynamical information of the forward shock. Furthermore, the SNR dynamics  is calculated \SCly\ with DSA including the feedback effects from NL-DSA on the SNR evolution. We include spatial profiles of the multi-wavelength emission which provide more stringent constraints on the parameters and lead to a better differentiation between the \pdecay\ and 
IC-dominated cases.

Within the approximations of our one-dimensional model, we find that leptonic models clearly fit the broadband observations better than \pdecay\ ones in a fashion strikingly similar to  broadband models of \SNRJ.\footnote{As in our models of \SNRJ, we only calculate emission from the shocked region between the \CD\ (CD) and FS and neglect emission from the \RS. Including additional thermal X-ray emission from the shocked ejecta would strengthen the case for a leptonic dominated model for the \gamray\ emission.} We emphasize that regardless of the approximations intrinsic to our model; our results clearly show that including a self-consistent calculation of the thermal X-ray emission is essential for any broadband model of a young SNR. For Vela Jr., as for \SNRJ, hadronic scenarios are excluded with high confidence in homogeneous models.

This paper is organized as follows: The first section provides a brief description of the latest version of the \crhydro\  simulation code we use as our modeling platform; the second section describes our models for Vela Jr.~and the various constrains from observation data; and in the last part, we discuss the implication of our models on our understanding of the environment of Vela Jr., the acceleration of CR particles at the blast wave, and the possible production mechanism(s) of the observed broadband emission from radio to the TeV band. Some concluding remarks follow.    

\section{The \textit{CR-hydro-NEI} Simulation Code}

When it comes to modeling emission from young shell-type SNRs with strong forward shocks, it is very important to take into account the non-linear aspects of DSA and its coupling to the hydrodynamics and plasma conditions in the shocked gas. 
The high Mach number shocks in young SNRs are expected to be efficient particle accelerators and with efficient DSA, the CR production, shock structure, shocked gas temperature, and magnetic fields are coupled and feedback effects cannot be ignored.
Here we employ a \crhydro\ code which can effectively model young, shell-type SNRs like Vela Jr.~with these non-linear effects accounted for self-consistently. In this section, we briefly describe a generalized version of the \crhydro\ code that includes several updates and additional functions not included in the description given in \citet{LEN2012}. 

\subsection{Coupled SNR hydrodynamics, NL-DSA, and NEI}
 
The development of the \crhydro\ code used here has been described in detail in a number of recent papers \citep[e.g.,][]{EPSBG2007,PES2009,PSRE2010,EPSR2010,ESPB2012}, with the most recent ``generalized" version given in \citet{LEN2012}. Briefly, the SNR hydrodynamics are modeled with a one-dimensional hydro simulation based on VH-1 \citep[e.g.,][]{BE2001}.
The SNR simulation provides the evolving shock speed, sonic and \alf\ Mach numbers, and other quantities necessary for the shock acceleration calculation. The NL-DSA calculation is done with a \SA\ solution based largely on the work of P. Blasi and co-workers 
\citep[e.g.,][]{BGV2005,CBAV2009} \citep[see][for a full list of references]{LEN2012}.
With suitable parameters, the \SA\ calculation provides the shock compression ratio, amplified magnetic field, and full 
proton and electron spectra which
are used to calculate the broadband continuum  radiation from \syn, \brem, IC, and \pdecay\ emission.

The SNR evolution is coupled to the CR production by modifying the  
standard hydrodynamic equations  through a change in the equation of state from the influence of \rel\ CRs, energy loss from escaping CRs, and magnetic pressure.\footnote{We note that only a ``scalar" magnetic pressure is used in our 1D hydro simulation.}
The changes produced in the hydrodynamics by NL-DSA also modify the evolution and ionization state of the shocked thermal plasma and these changes are \SCly\ included in our \NEI\ (NEI) calculation of the thermal X-ray emission. 

We thus obtain a model of an evolving SNR where the non-thermal continuum and thermal line emission are calculated \SCly. Treating thermal and non-thermal processes consistently is required when CR production is efficient because the production of \rel\ CRs can strongly influence the density and temperature of the shock-heated thermal plasma \citep[e.g.,][]{Ellison2000}. In such a self-consistent calculation, it is not possible, for instance, to adjust parameters so that \pdecay\ matches GeV-TeV observations without modifying the fit to thermal X-ray line emission. As we have demonstrated in our fits to \SNRJ\ 
\citep[e.g.,][]{ESPB2012,LEN2012}, this coupling strongly constrains broadband emission models.

\subsection{Precursor CR populations and emission}

To determine the CR proton distribution function $f^\mrm{cr}(x,p)$ at a certain position $x$ in the precursor 
upstream of the forward shock, we follow the recipe described in \citet{LEN2012}  [see equation~(17)] and references therein. 
As for the electrons, we implicitly assume that they do not affect the shock structure and dynamics due to their low energy density compared to the protons, which we have checked to be the case a posteriori for all models presented here. We also assume that radiative cooling of the electrons occurs mainly in the downstream region where the B-field is highest and the electrons spend most of their time scattering per shock-crossing. As a result, this allows us to calculate the electron distribution in the precursor easily using the precursor profile and the distribution function at the subshock in the same way as for the protons \textit{after} the DSA solution has been obtained at every time step. 

At each position $x$ in the precursor, we read off the local environment variables such as the gas density, the B-field strength and the temperature as provided by the DSA solution to calculate the local photon emissivity. The total emission spectrum from the CR precursor is thus obtained by an integration over the precursor volume.

\subsection{Secondary particle production and emission in the SNR shell}
Accelerated protons and heavier ions will interact with the background gas and produce charged pions through p-p interactions. These pions further decay into secondary particles such as e$^+$, e$^\mrm{-}$ and $\nu$'s. The charged secondaries can accumulate in the SNR shell and radiate as the shock propagates into the upstream medium. We have implemented this process in the \crhydro\ code by following the continuous production of the secondaries in the 
post-shock region, as well as the adiabatic and radiative losses of the \pairs. We only consider secondary production and radiation in the post-shock region because the density and magnetic field are considerably higher than in the shock precursor.
The relevant inclusive secondary cross-sections are obtained using the parameterized model by \citet{Kamae06}, as we do for the gamma-rays. Contribution from 
heavy ions like p-He is taken into account via an effective enhancement factor.  These secondaries contribute to the photon emissivity via synchrotron, IC and 
bremsstrahlung and is estimated at the end of a simulation.  

We note that there is a possibility that these secondary 
\pairs\ can be injected into DSA and be `reaccelerated' in the same manner as the primary electrons. This is especially true for those being generated in the CR precursor. However, since most of the secondary particles are produced and accumulated in the DS region, and the probability of these populations traveling back to the shock from their production sites in order to be injected into acceleration is relatively low, we will ignore the process of reacceleration. 
We will show that, whilst the secondary particles do not contribute significantly to the volume-integrated broadband photon spectrum for all models presented in this work, the production of these particles in the shocked gas and their accumulation can have important effects on the spatial variation of synchrotron emission in some specific X-ray bands for a hadronic model. This result will be discussed in detail in Section~\ref{section:xray}.

\subsection{Multi-wavelength emission profiles}

Using the spatial information from the \crhydro\ code, it is relatively straightforward to obtain model predictions for multi-wavelength emission profiles. Line-of-sight projection effects are calculated assuming spherical symmetry. The projected radial (i.e., outward from the SNR center) profiles, integrated over a set of chosen wavebands, are convolved with gaussian kernels to mimic the 
point-spread-functions of various instruments and facilitate direct comparison with data.    

\section{Models}
\label{sec:models}
The radio-to-TeV spectral and morphological features of Vela Jr.~largely resemble SNR {\SNRJm}. The observed similarities between Vela Jr.~and \SNRJ\ include: (i) relatively dim radio emission with a flat spectral index; 
(ii) radio and X-ray emission which are predominately synchrotron in origin with ring-like morphologies, and with no evidence of thermal emission from either the shell or the ejecta (central) region; (iii) luminous in TeV energies, but relatively faint in the GeV band when compared to other gamma-ray SNRs, especially the middle-aged remnants like W51C, W44, and IC 443 \citep[e.g.,][]{AbdoEtalW51C2009, AbdoEtalW442010, AbdoEtalIC4432010}; and (iv) TeV emission is spatially resolved to a shell-like structure and correlates well with the X-ray and radio images. 
As might be expected, the observational similarities between Vela Jr.~and \SNRJ\ lead naturally to a similar conclusion from our 
\crhydro\ model that IC dominates the GeV-TeV emission process.

\subsection{Input parameters}
We have constructed two representative models: a ``hadronic" model where the 
GeV-TeV emission is \pdecay\ dominated, and a ``leptonic" model  where it is IC dominated.
For both models we assume:
(i) the associated supernova (SN) had a kinetic energy of $\EnSN = 10^{51}$~erg;
(ii) the age of the remnant is 2500~yr to match the observed size of the SNR at the distances assumed in Table~\ref{table:param};
(iii) the injection parameter which determines the acceleration efficiency of DSA is set at 
$\Inj \equiv p_\mrm{inj}/p_\mrm{th} + u_2/c = 3.6$, corresponding to a fraction $\eta \sim 2-3 \times 10^{-4}$ of thermal particles being injected in DSA at any time 
\citep[note that our $\Inj$ is $\xi$ in][]{BGV2005};
(iv) we have conservatively assumed that the electron temperature equilibrates with the proton temperature through Coulomb collisions unless otherwise mentioned; 
(v) the spectrum of the background photon fields that are up-scattered by the relativistic electrons to gamma-ray energies (IC) is taken from \citet{PMS2006} at the position of the SNR, which contains the cosmic microwave background (CMB) and the local interstellar radiation fields in the infrared and optical bands;
(vi) the upstream gas temperature is fixed at $T_0 = 10^4$\,K regardless of the gas density gradient; 
(vii) a number fraction of helium $\fHe=0.0977$ is specified for the plasma, which contributes additionally to the various photon emission. 
However, other than this scaling, we ignore the acceleration of heavy ions in this paper;  
(viii) we assume that the ambient B-field is oriented quasi-parallel to the shock normal over the whole surface, that is, we assume that the magnetic field geometry is unimportant as would be the case for efficient DSA where the self-generated magnetic turbulence mediating DSA is totally tangled in the CR precursor and downstream from the shock, i.e., the Bohm limit for CR diffusion is obtained. 
For full details of the formulation, see \citet{LEN2012}. 

An important difference between our hadronic and leptonic models is that the hadronic model evolves in a uniform upstream medium, typical of a Type Ia SN, and the leptonic model evolves in a wind cavity, typical of a core-collapse SN. 
While the observational evidence favors a \CC\ origin for 
the Vela Jr.~SNR, we could not find a set of \CC\ parameters that would produce a reasonable fit to the broadband data for a hadronic model radiating in a pre-SN wind. This is because the hadronic model needs a high ambient gas density and high B-field strength in order to simultaneously explain the 
radio/X-ray synchrotron emission and the gamma-ray flux. We thus used a uniform ISM model for our hadronic fit. On the other hand, a leptonic model requires a lower ambient gas density to suppress the \pdecay\ gamma-rays, and this is natually realized by a wind bubble environment for a remnant sitting close to the Galactic plane. 
More details will be given in Section~\ref{section: results} where we will further discuss this point. Consistent with the different SN scenarios, we assume an ejecta mass $\Mej=1.4\,\Msun$ for our hadronic model and $\Mej=3\,\Msun$  for our leptonic model.

In Table~\ref{table:param}, we summarize the main input parameters and output quantities measured at the end of the simulation for each model. We constrain the input parameters via a satisfactory fit to the observed broadband spectrum from radio to TeV energies, as well as a general agreement  of the dynamical variables (e.g., radius of the forward shock) with current observations. The distance to the SNR is recently constrained by \citet{Katsuda2008} using proper-motion measurements over a 7-year time period. They provided the estimate equivalent to $d_\mrm{SNR} \gtrsim (750 \pm 205) \times (v_\mrm{sk}/3000\ \mrm{km}\ \mrm{s}^{-1})$~pc. We will adopt $d_\mrm{SNR}$ for our models which is consistent with this limit. 

\section{Results and Discussion}
\label{section: results}
\subsection{Dynamics}
%
\begin{figure}
\centering
\includegraphics[width=9cm]{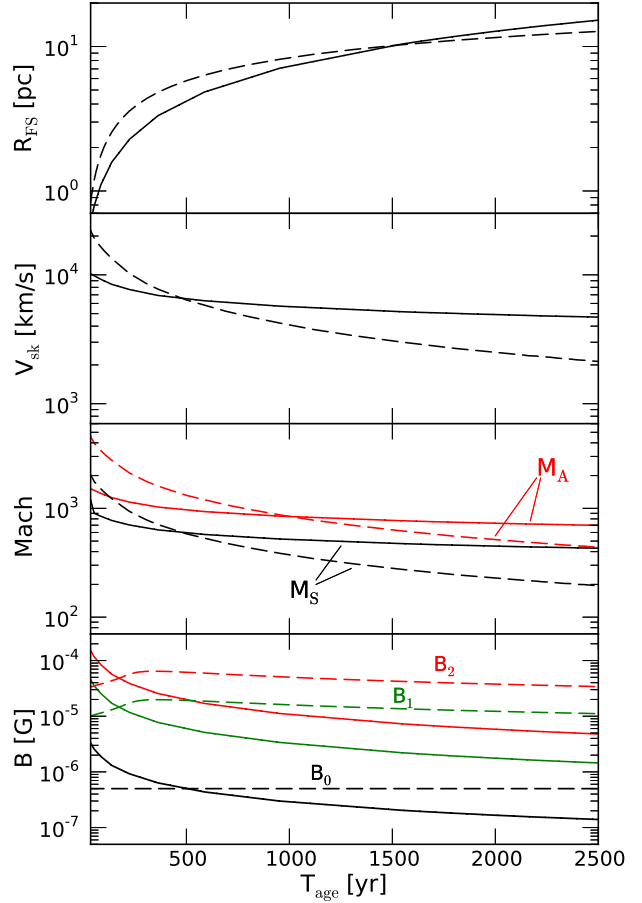}  
\caption{
Time evolution of output quantities for our leptonic model (solid lines) and hadronic model (dashed lines). From the top panel: (1) shock radius $R_\mrm{FS}$; (2) shock velocity $V_\mrm{sk}$; (3) sonic Mach number $M_\mrm{S}$ (black) and Alfv\'{e}n Mach number $M_\mrm{A}$ (red) of the forward shock; and (4)  B-field strength in the far upstream $B_0$ (black), right in front of the subshock $B_1$ (green), and immediately behind it $B_2$ (red).
}
\label{fig:model_evo_1}
\end{figure}
%
\begin{figure}
\centering
\includegraphics[width=9cm]{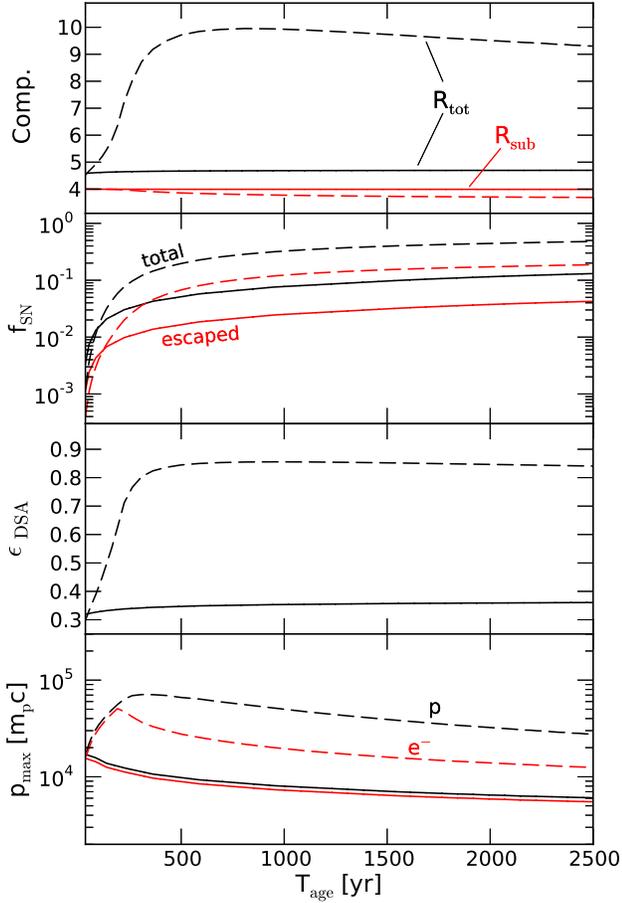}  
\caption{Continuation of Figure~\ref{fig:model_evo_1} where again, solid curves are our leptonic model and dashed curves are our hadronic model. From the top panel: (1) compression ratios at the subshock $R_\mrm{sub}$ (red) and total $R_\mrm{tot}$ (black); (2) fraction of $\EnSN$ converted into relativistic particles $f_\mrm{SN}$ for the total accelerated (black) and escaping (red) populations; (3) acceleration efficiency $\epsilon _\mrm{DSA}$; and (4) maximum momentum, $p_\mrm{max}$, attained by the accelerated protons (black) and electrons (red).   }
\label{fig:model_evo_2}
\end{figure}

We plot the time evolution of selected important physical quantities for the hadronic and leptonic models in Figures~\ref{fig:model_evo_1} and 
\ref{fig:model_evo_2}.
The very different undisturbed upstream environments give rise to large differences in the dynamical behavior of the two models. 
With the lower B-field and gas density in the wind cavity for the leptonic model, the following are expected and observed compared to the 
hadronic model: (1) the shock sweeps up material at a much lower rate, so that the shock speed and Mach numbers decay slower, and shock radius increases faster, 
with time; (2) the acceleration efficiency, and hence the fraction, $\fSN$, of supernova explosion energy, $\EnSN$, converted into CR particles at the current age is much lower 
($\fSN \simeq 0.14$ vs 0.48) due 
to the slower injection rate of thermal particles from the shocked gas into DSA. This is also reflected in the much more moderate shock modification 
(i.e., smaller $R_\mrm{tot}$) than for the hadronic model on average. The high $\fSN$ fraction of the hadronic model is consistent with the result of \citet{Tanaka2011}; (3) the lower B-field in the DS region ($B_2$) results in a longer synchrotron loss time-scale, in 
this case longer than the acceleration time-scale throughout the age of 2500 yr, resulting in a common $p_\mrm{max}$ for electrons and protons. On the contrary, the 
hadronic model has a much higher ambient B-field strength 
in the uniform ISM,  such that $p_\mrm{max}$ of electrons becomes limited by the synchrotron loss (and IC loss to a lesser extent) starting from about 
200~yr. These evolutionary differences relate directly to the broadband emission of the SNR, as we will discuss in the upcoming sections.  

\subsection{Broadband Photon Spectrum}
%
\begin{figure}
\centering
\includegraphics[width=9cm]{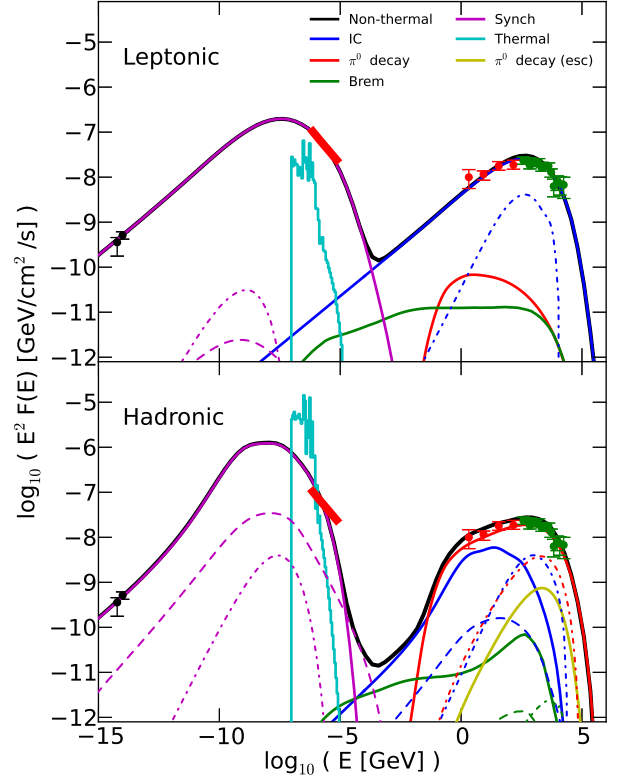}  
\caption{Lower panel: Broadband SED fit with parameters tuned so the gamma-rays are dominated by  hadronic emission. The solid lines 
show emission spectra produced by the primary CRs in the post-shock region, with the exception of the yellow line which shows emission from the interaction of the escaped CRs with the upstream medium beyond the free escape boundary (FEB). The dashed lines show emission in the post-shock region produced by the secondary e$^+$ and e$^\mrm{-}$ originating from p-p interactions and the subsequent decay of charged pions. The dash-dotted lines show emission from the CR precursor region ahead of the forward shock. 
Upper panel: Same as the lower panel but with parameters tuned 
so that the gamma-rays are dominated by IC emission. 
All model spectra in the lower panel are boosted by a factor of $\sim 5$ to match the observed fluxes.
(Data: black - radio \citep{Combi1999}; red band - \textit{ASCA} GIS \citep{Aharonian2007b}; 
red - \textit{Fermi} LAT \citep{Tanaka2011};  green - \textit{H.E.S.S.} \citep{Aharonian2007b}.
}
\label{fig:model_spec}
\end{figure}

As mentioned in Section~\ref{sec:models}, to obtain 
reasonable fits to the observed photon spectra, the two models require that the SNR evolves within very different surrounding environments. 
For the hadronic model, in order for the \pdecay\ photons to dominate over the IC component in the gamma-ray band, the gas density has to be high and the number ratio between electrons and protons ($\Kep$) has to be particularly low ($\Kep \sim 10^{-4}$). The high density requirement argues against the SNR propagating into a pre-SN wind with rapidly decreasing density and we are forced to assume the blast wave runs into an interstellar medium (ISM) with uniform gas density and magnetic field strength. 

In constrast, the leptonic model requires that the forward shock propagates into a magnetized pre-SN wind region. This introduces a descending gradient for the gas density and B-field. The low shocked gas density in the downstream photon-emitting region suppresses  the \pdecay\ and \brems, as well as the X-ray line emission, relative to the IC component. 

Figure~\ref{fig:model_spec} shows the corresponding fits of the broadband spectra to currently available observations.
We see that both models can explain the radio and GeV-to-TeV gamma-ray data reasonably well under the respective choices of parameters. 
In the X-ray band, however, the hadronic model has difficulty reproducing the X-ray photon index observed by \textit{ASCA}, mainly due to the much faster synchrotron loss rate caused by the higher B-field in the hadronic model.
The higher ambient gas density required for \pdecay\ dominance in the gamma-ray band is also accompanied by a high flux from thermal X-ray lines, which exceeds the X-ray flux observed by \textit{ASCA}
in the soft X-ray energy range below a few keV, despite the adoption of a conservative model of electron heating via Coulomb collisions in the post-shock region. 
The \SCly\ calculated thermal X-ray emission provides a strong additional model constraint.

Another important difference between the two models is the overall normalization factor required for the broadband emission to match the observed flux. For the leptonic model this factor is close to unity, but the hadronic model has to be boosted by a factor $\sim 5$ to explain the observed flux level.
%
%
This factor $\sim 5$ is included in all of the model spectra  shown in the lower panel of 
Figure~\ref{fig:model_spec}. 

\subsection{X-ray Spectrum and Its Spatial Properties}
\label{section:xray}
%
\begin{figure}
\centering
\includegraphics[width=8cm]{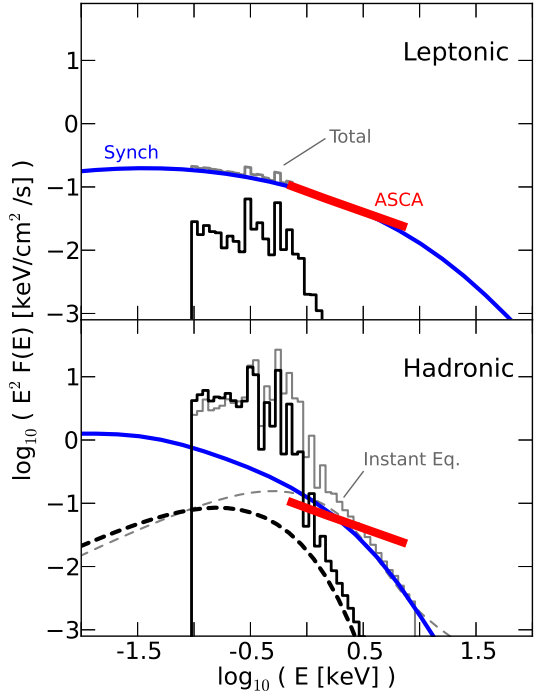}  
\caption{
X-ray spectra for our leptonic model (upper panel) and hadronic model (lower panel). In both panels, the blue and black solid lines correspond to the synchrotron (including secondary contribution if any) and thermal line emission respectively, while the dashed lines are the thermal continuum spectra. The red band is from \textit{ASCA} GIS observation. In the lower panel for the hadronic model,  the light grey lines also show the thermal spectra corresponding to a model with instantaneous equilibration between the electron and proton temperatures right behind the subshock, but is otherwise the same. In the top panel for the leptonic model, the grey line shows the total spectrum. 
As in Figure~\ref{fig:model_spec}, the hadronic model spectra are multiplied by $\sim 5$ to match the observed broadband flux level. 
}
\label{fig:model_xray}
\end{figure}
A more in-depth comparison with the currently available X-ray data is valuable to differentiate SNR models. 
Firstly, to see the general agreement of the models with data, we plot the calculated non-thermal and thermal X-ray spectra against the \textit{ASCA} result \citep[e.g.,][]{Slane2001, Aharonian2007b} in 
Figure~\ref{fig:model_xray}. For the hadronic model, to show the possible range of variance of the thermal spectrum based on different equilibration models for the electron temperature, we also show the calculated thermal spectrum assuming that the electron temperature equilibrates instantaneously with the proton temperature behind the FS. Coulomb equilibration and instant equilibration are two extremes for electron heating and the actual heating is very likely to lie somewhere in between.
For either extreme in heating, the hadronic model predicts a thermal X-ray flux substantially higher than what observation implies (especially below $\sim 2$~keV), in addition to its failure to reproduce the non-thermal spectral index. 
The leptonic model, on the other hand, shows a general good fit to the data without an 
over-prediction of the thermal emission. 

In this regard, we notice that \citet{Berezhko2009} have proposed a hadronic model in which the shock is running into a wind bubble region created by the progenitor star embedded inside a high density gas cloud. We do not consider such a situation in this work since, in such a high density environment 
($n_\mrm{gas} \gg 0.01$~cm$^{-3}$), the resultant thermal X-ray lines are expected to violate the observation data even more than is the case for our hadronic model.

\begin{figure}
\centering
\includegraphics[width=8.5cm]{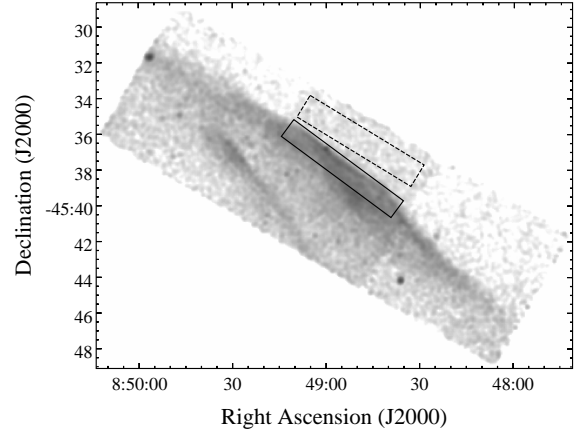}  
\caption{
\textit{Chandra} image of the NW region of Vela
Jr.  The regions used for source and background spectra are shown
by the solid and dashed boxes respectively. A circle containing a
faint point source falling within the source region was excluded
in the spectral extraction.
}
\label{fig:chandra_image}
\end{figure}

Secondly, we consider the X-ray spectrum near the bright NW rim of
the remnant.  Vela Jr.~was observed with the \textit{Chandra X-ray
Observatory} on 2003 January 5 and 6 (ObsIDs 3446 and 4414) using
the Advanced CCD imaging Spectrometer (ACIS). Standard cleaning and
data reduction were performed using CIAO version 4.4.  The merged
observations yielded a total exposure time of 73.9 ks.  Spectra,
extracted using the {\sl specextract} software, were obtained from
a narrow region encompassing the forward shock of the SNR (solid
box in Figure~\ref{fig:chandra_image}) and from a background region (dashed box) used to
account for projected emission from the Vela SNR as well as internal,
Galactic, and extragalactic background emission.

The predicted thermal and nonthermal emission from \textit{CR-hydro-NEI} 
for the projected spectral region indicated was converted into a
table model for use in {\sl xspec} for both the hadronic and leptonic
models. The {\sl tbabs} model for interstellar absorption was
applied, and the model was convolved with the instrument response
function and fit to the data. To account for possible excess
foreground emission from the Vela SNR, a thermal nonequilibrium
ionization model ({\sl nei}) with a distinct {\sl tbabs} absorption
component was included.

\begin{figure}
\centering
\includegraphics[width=9cm]{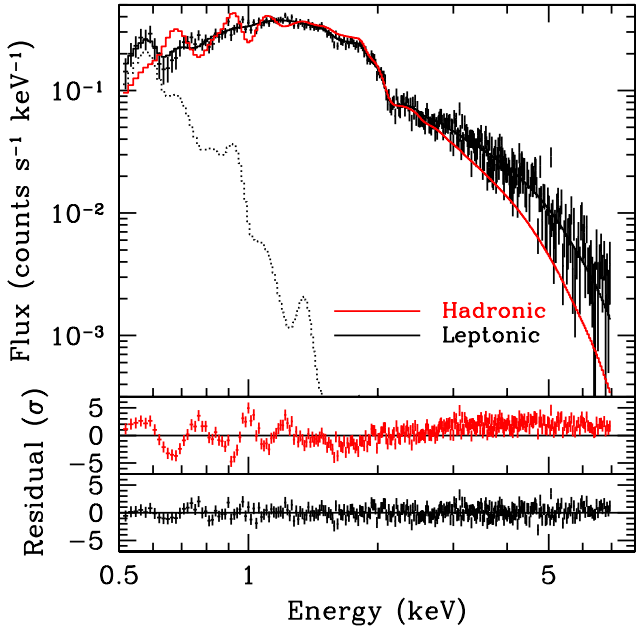}  
\caption{
\textit{Chandra} spectrum of the NW rim compared
with the spectra from our leptonic (black) and hadronic (red) models.
The model spectra are folded with the instrument response function
and the fits reveal an interstellar absorption with a column density
of $n_\mrm{H} \approx 4 \times 10^{21}$~cm$^2$. An additional thermal
component (dashed histogram) with $T_e \sim 0.3$~keV and $n_\mrm{H} \sim 2 \times
10^{21}$~cm$^2$ is added to the leptonic model to explain the background residual at low
energies. 
}
\label{fig:model_chandra_spec}
\end{figure}

The best-fit results are shown in Figure~\ref{fig:model_chandra_spec}, where the leptonic model
fit is shown as a black histogram, and the hadronic model is shown
in red.  The dashed histogram corresponds to the 
additional {\sl tbabs}$\times${\sl nei} model.  It is clear that the hadronic model
provides a poor representation of the X-ray data, both at low
energies, where the predicted line emission is not observed, and
at high energies, where the model is much steeper than the observed
spectrum. The leptonic model provides an excellent fit to the data
($\chi_\nu^2 = 1.05$ for 327 degrees of freedom). The column density
is $n_H = (4.3 \pm 0.2) \times 10^{21}{\rm\ cm}^{-2}$, in good
agreement with previous measurements \citep{Slane2001, Pannuti2010}
and the accompanying thermal component has a temperature
of $0.26^{+0.32}_{-0.08}$~keV with a column density of $(0 - 3.3)
\times 10^{21}{\rm\ cm}^{-2}$, consistent with observed properties
for soft thermal emission from the Vela SNR.  The best-fit normalization
for the \textit{CR-hydro-NEI} component is about 15\% higher than the
ratio of the spectral extraction region length to the entire SNR
circumference, which is reasonable given that the rim is somewhat
brighter at this position (see Figure~\ref{fig:chandra_image}).

%

\begin{figure}
\centering
\includegraphics[width=9cm]{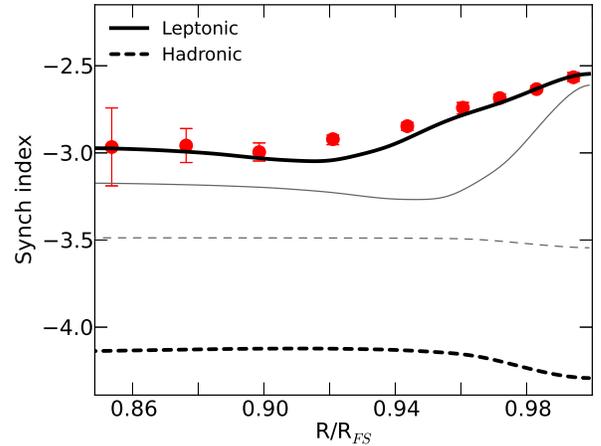}  
\caption{
Line-of-sight dependence of the non-thermal X-ray spectral index. The synchrotron indices are extracted from regions with various radial distances from the SNR center and in the energy range of $2 - 10$~keV. The black thick lines are the result from our best-fit models. 
Red points are spectral fitting results from \citet{KHU2013} using archival \textit{XMM Newton} data in the same energy range. 
The thin grey lines correspond to the modified models described in Section~\ref{section:sensitivity} below.
Note that the contact discontinuity lies at $\mrm{R}/\mrm{R}_\mrm{FS} = 0.81 - 0.82$ for both hadronic and leptonic models.
}
\label{fig:synch_index}
\end{figure}

Thanks to the high angular resolution of \textit{Chandra}, it is also possible to study spatial variation of the spectral properties near the FS \citep[see, e.g.,][]{Pannuti2010}. Correspondingly, we calculate the synchrotron photon index for a series of line-of-sight projected regions from the models which have different radial distances from the FS. The results are shown in Figure~\ref{fig:synch_index}. Besides the absolute difference of the index values, the hadronic and leptonic models show very distinct trends of the index variation as a function of distance from the FS. For our leptonic model, the synchrotron spectrum softens as we look inward from the FS, whereas our hadronic model exhibits a much weaker dependence on position. 

The major factors which determine the spectral index of the 
non-thermal X-rays at a certain distance from the SNR center are the following: (1) Energy loss history of the primary accelerated electrons, which is critical to the high-energy cut-off of the primary synchrotron component; and (2) the relative importance of the secondary \pairs\ to the primary electrons at that position. These secondaries are produced by the trapped accelerated protons interacting with the shocked gas, which lose energy radiatively like the primaries and meanwhile accumulate in number with time. 
Different from the primaries, which are accelerated by the shock and then advected downstream without replenishment, the secondaries at each position in the shocked gas are continuously generated by the advected high-energy protons whose spectra do not evolve with time significantly other than from adiabatic losses. The spectral shapes of the secondary and primary synchrotron components are hence typically different.
%
For our leptonic model, the effect from secondaries is obviously small due to the very low ambient gas density in the wind bubble. As a result, the energy loss of the primaries dominates the spatial behavior of the synchrotron radiation. Since primary electrons accelerated by the shock at an earlier phase and now residing further inward in the SNR shell have suffered from synchrotron/IC and adiabatic losses for a longer period of time than those just being freshly accelerated, the cutoffs of their spectra are found at lower energies. In addition, due to the falling B-field strength with radius in our wind model, those primary electrons accelerated earlier have experienced a higher synchrotron loss rate on average than those residing closer to the FS at the current age.
These lead to the softening of the synchrotron spectrum as shown by our result.

For the hadronic model, the secondaries cannot be neglected.
In the volume integrated photon spectrum (Figure~\ref{fig:model_spec}), the primary component dominates over the secondary one until around 10~keV. However, by looking deeper towards the SNR center where the number of secondary particles is larger due to accumulation and the primaries have a lower cut-off energy from a longer period of radiative loss, the secondary spectrum can dominate over the primary spectrum at a few keV and harden the resultant total spectrum in the $1 - 5$~keV energy range. 
We find that this effect can compensate the energy loss of the primaries in the SNR shell and, for the relatively moderate ambient gas density we assume, lead to a weak variation of the synchrotron index with radius after projecting our spherically symmetric model along the line-of-sight. 
With a higher ambient gas density, a spectral hardening can be expected as we look inward from the shock. Using \textit{Chandra} and \textit{XMM Newton} data respectively, \citet{Pannuti2010} and \citet{KHU2013} have recently discovered a trend of softening for the non-thermal spectrum as the line-of-sight is moved inward from the FS towards the SNR center. This trend provides further support for our leptonic scenario. For a direct comparison, we overlaid in Figure~\ref{fig:synch_index} the fitted photon indices given by \citet{KHU2013} and found a reasonable agreement with our result. Additionally, the phenomenological spectral evolution model invoked by \citet{KHU2013} shows that the typical downstream B-field should be around a few $\mu$G in order to explain the observed softening trend, which is highly consistent with the field strength obtained in our leptonic model.

\subsection{Gamma-ray Morphology}

\begin{figure}
\includegraphics[width=9cm]{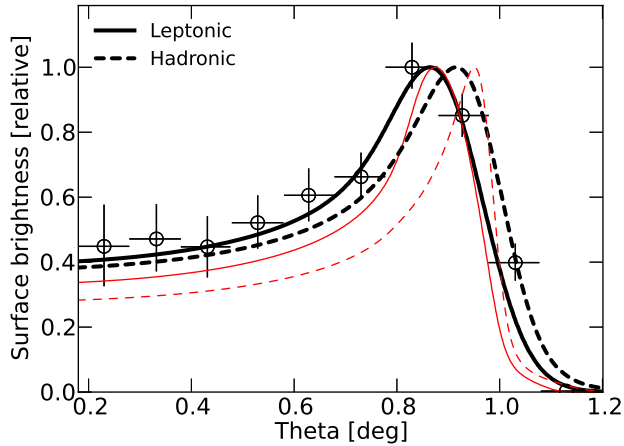} 
\caption{Radial profile of gamma-ray emission (300~GeV - 20~TeV) compared with \HESS\ measurements of the northern shell. The black thick curves are model profiles smoothed by a Gaussian kernel with $\sigma = 0.06^\circ$ corresponding roughly to the PSF of the \HESS\ observation. The red thin curves are the model profiles expected under the advertised best angular resolution of \textit{CTA} ($0.02^\circ$).
}
\label{fig:TeV_profile}
\end{figure}

The current generation of 
ground-based Imaging Atmospheric Cherenkov Telescopes (IACTs) possesses sufficient angular resolution to discern the spatial structures of SNRs with large angular sizes in sub-TeV to multiple TeV energies. Recent \HESS\ observations of Vela Jr.~\citep{Aharonian2007b} have measured the radial surface brightness profiles of its shell-like emission from 300~GeV to 20~TeV with a resolution of about $0.06^\circ$. We compare our model profiles with the \HESS\  measurements for the brightest northern part in 
Figure~\ref{fig:TeV_profile}. The profiles are extracted from the model gamma-ray image after convolution with the PSF of the observation. The results show that both models predict gamma-ray emission profiles reasonably compatible with the data, and unlike the X-ray result, closely resemble each other after PSF smoothing. 

The reason behind this relative similarity in the gamma-ray band can be explained by the following: X-ray synchrotron emission produced by relativistic electrons is highly sensitive to the radiative loss rate. %
Electrons experience 
much faster energy loss in our hadronic model than the leptonic model, as mentioned above, which results into a sharp X-ray profile close to the FS for the hadronic case, and a more spread out profile behind the FS for the leptonic case. 
In the gamma-ray band, however, radiative loss plays a much less important role.
In the hadronic model,  
the gamma-ray photons mainly originate from CR protons where losses can be ignored,
while in the leptonic model the electrons responsible for the IC emission are only subject to minor losses due to the low magnetic fields.
%

Nevertheless, the gamma-rays originate from totally different mechanisms for the hadronic and leptonic cases
so differences are predicted for more precise measurements.
Future ground-based gamma-ray observatories, such as the upcoming Cherenkov Telescope Array (\textit{CTA}), 
will be able to differentiate the models with unprecedented spatial resolving power and sensitivity. For illustration, we 
include in Figure~\ref{fig:TeV_profile} our
model profiles using the advertised best angular resolution of \textit{CTA} (about $0.02^\circ$). The red curves imply that future gamma-ray observations may be able to differentiate between the predicted
radial profiles of Vela Jr.

\subsection{Possible Thermal X-ray Line Detection by \textit{Astro-H}}

\begin{figure}
\centering
\includegraphics[width=9.5cm]{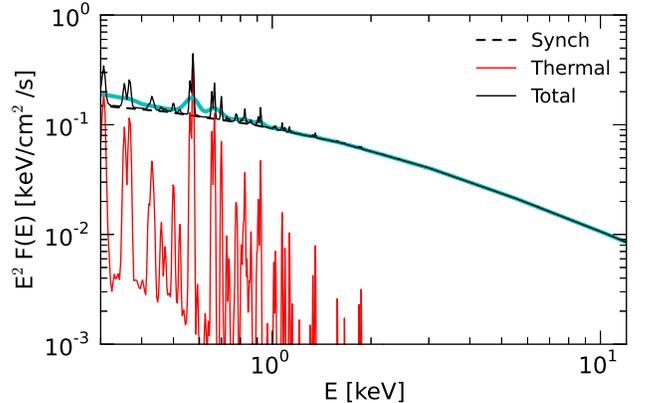}  
\caption{Thermal emission lines and non-thermal spectrum from 0.3 to 12~keV from the leptonic model. The model spectrum is filtered with a Gaussian kernel with $\Delta$E (FWHM) = 7~eV consistent over the whole energy range to imitate the (minimum) target energy resolution of the SXS instrument onboard the next-generation X-ray observatory \textit{Astro-H}. A thick cyan line is overlaid to show the total spectrum under a fixed resolution of 60~eV for comparison.
}
\label{fig:spec_astro_h}
\end{figure}

TeV-bright SNRs like Vela Jr.~and RX J1713.7-3946 have shown no sign of strong thermal emission with current X-ray observations. However, with the advent of future instruments of superior spectral resolution, it is very possible that thermal X-ray lines will be discerned in the non-thermal dominated spectrum.  As a first look, we plot the calculated X-ray spectrum in Figure~\ref{fig:spec_astro_h} for our leptonic model in the energy range of 0.3 to 12~keV, the designed energy coverage of the \textit{Soft X-ray Spectrometer} (SXS) onboard the next-generation X-ray space observatory \textit{Astro-H}. The emission lines are filtered by a Gaussian kernel with a FWHM of 7~eV to mimic the target baseline spectral resolution of SXS.\footnote{We assume a pure spherical expansion of the SNR shell, such that near the FS (i.e., the SNR rim) the line broadening/shifting effect from expansion is negligible.} 

For comparison, we show the leptonic model spectrum with 
a resolution of 60~eV (a typical ballpark value for currently operating X-ray observatories). We see that using the resolving power of SXS, the stronger sub-keV lines can be recognized very visibly on top of the synchrotron spectrum, and it is very possible that they will be detected by \textit{Astro-H}. Detection of such lines will be important to further constrain our model parameters such as the local gas compositions and the ion temperatures and densities. In a follow-up paper in preparation, we will perform a more detailed study of the X-ray spectrum by performing formal \textit{Astro-H} SXS spectral simulations based on our models, and take into account thermal Doppler broadening of the emission lines.\footnote{For the typical post-shock temperature in our leptonic model, the broadening by thermal motion of the ions is around $1\%$, assuming that the temperature ratios among the ion species are mass-proportional. Therefore, roughly speaking, SXS with a FWHM of at most 7~eV will be able to identify thermal broadening of the stronger lines above approximately 0.7~keV, where there are a few. We will perform a more detailed calculation of the ion temperatures in our follow-up paper to better assess this possibility.}    
%

\subsection{Sensitivity of Results on Model Parameters} 
\label{section:sensitivity}
We emphasize that while small variations in parameters in either of our ``best-fit" models do not strongly modify the results for that model, 
the parameter sets for the two scenarios are well separated and it is not possible to go smoothly from one set to the other while maintaining a satisfactory fit to the observations. 

To illustration this point and see how our models respond to parameter changes, we create modified versions for each of our models in which we purposely tune a set of important parameters so that it is positioned closer to the parameter space of the competing best-fit model. Two of the most important parameters that distinguish our leptonic and hadronic scenarios include the electron-to-proton number ratio at relativistic energies $K_\mrm{ep}$, and the upstream gas density $n_0$ (or the mass loss rate $\dMdt$ and wind speed $\Vwind$ when a pre-SN wind is present). For the leptonic case with a massive star as progenitor and a wind cavity, by recalling that the IC flux $f_\mrm{IC} \propto n_0K_\mrm{ep}$ \citep[e.g.][]{EPSR2010},  we explore an altered model where $K_\mrm{ep}$ is lowered by a factor of 5 (i.e. $3 \times 10^{-3}$) and the wind matter density increased by the same factor to roughly maintain the level of $f_\mrm{IC}$.\footnote{Only decreasing $K_\mrm{ep}$ by such a factor will not visibly change the broadband spectral shape, but will require an overall normalization factor substantially larger than 1, thus excluding such an option.} The wind magnetization is fixed so as to keep the synchrotron-to-IC flux ratio $f_\mrm{syn}/f_\mrm{IC}$ roughly unchanged. On the other hand,  we explore an altered model for the hadronic case where $K_\mrm{ep}$ is boosted up by a factor of 5 while $n_0$ is decreased by a factor of 1.7, so that the model is now leaning toward the leptonic parameter space. Note that since the \pdecay\ flux $f_{\pi^0} \propto n_0^2$, this will decrease $f_{\pi^0}$ and increase $f_\mrm{IC}$ by roughly the same factor. For both cases, we also adjust the assumed age of the remnant, within the allowable range of $1700 - 4000$~yr implied by observations, from the default age of 2500~yr in order to achieve the same SNR angular size as the original model. Other than what described above, all other parameters are kept fixed.

\begin{figure}
\centering
\includegraphics[width=9cm]{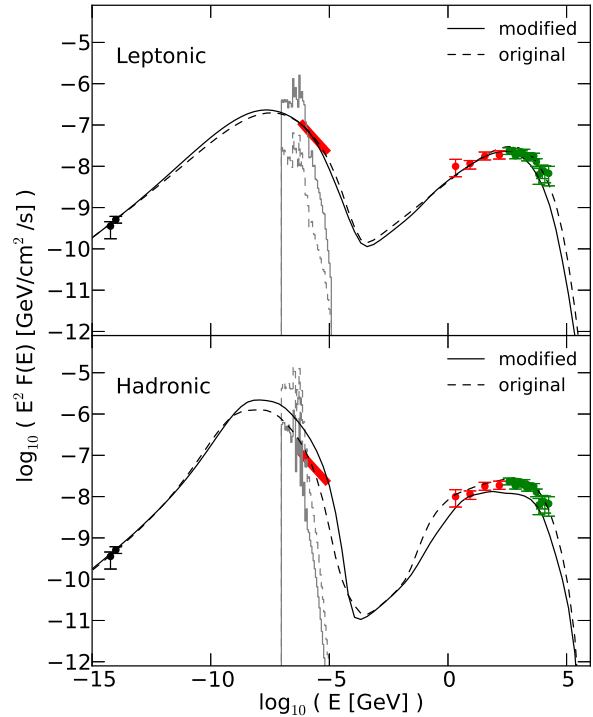} 
\caption{Comparison of the broadband spectra of our original best-fit models (dashed lines) with their respective modified versions (solid lines) to show the sensitivity of our models on parameter choices. In each panel, the black lines show the total non-thermal spectrum, and the grey lines show the accompanying thermal X-ray line emission. The data points are the same as Figure~\ref{fig:model_spec}. See text for details about the modified models.
}
\label{fig:sed_sensitivity}
\end{figure}

The effect on the total broadband spectrum is shown in Figure~\ref{fig:sed_sensitivity}. We can see that the modified leptonic model produces a non-thermal spectrum with an acceptable fit to the data except the highest energy TeV points and a X-ray spectral index a little too soft. But most importantly, it already starts to face the difficulty of over-predicting the thermal emission in the X-ray band due to the enhanced ambient gas density, which is the same problem encountered by our hadronic model. Another surfacing problem not shown in this plot is that the spatial dependence of the synchrotron index now fails to explain the \textit{XMM Newton} data by showing a softening trend too strong towards the SNR center (see Figure~\ref{fig:synch_index}). For the modified hadronic model, although the thermal X-ray reduces back to a flux level more compatible with data than the original model, the broadband spectral fit becomes unacceptable by failing to explain the non-thermal X-ray flux level simultaneously with the gamma-rays, due to the enhanced leptonic contributions which worsen the fit. To achieve a good fit again, we find that a much lower B-field must accompany the increased $K_\mrm{ep}$ and decreased $n_0$, which is simply what being realized in the wind model we are employing for the leptonic model. The conclusion from this analysis is hence that it is impossible to maintain a good fit to data by continuously tuning some influential parameters from the leptonic to the hadronic case (and vice versa) unless a drastic change of the underlying SNR environment is involved, i.e.~a rarified wind cavity against a uniform ISM model for the ambient medium.


\section{Conclusions}
\label{conclude}
We have investigated the TeV bright SNR Vela Jr.~with
a comprehensive modeling of its multi-wavelength emission 
from the radio to TeV band. 
Our one-dimensional, spherically symmetric NL-DSA model indicates that the SNR originated from the core-collapse explosion of a massive star inside a wind cavity with rarified gas density and magnetic fields. Based on the broadband continuum emission, with the added constraint from the \SCly\ determined thermal X-ray emission, we show that the GeV-TeV gamma-rays are predominately produced by IC emission from shock accelerated electrons with a much smaller  hadronic contribution from shock accelerated ions. Despite the fact that the leptons radiate more efficiently, here, as in all \SC\ DSA models of SNRs we are aware of, far more energy is put into \rel\ ions than electrons.
In summary, these conclusions are supported by the following results:

\begin{itemize}
\item In order for \pdecay\ gamma-rays from shock accelerated ions to dominate IC emission from electrons, the ambient pre-shock density must be above some limit. Despite a wide range in possible thermal equilibration models for electron heating and the \NEI\ of the shocked plasma, the  density required for \pdecay\ dominance is accompanied by strong thermal line emission in the X-ray band which is well 
above \textit{ASCA GIS} observations (see the lower panel of 
Figure~\ref{fig:model_spec}).
This conclusion is similar to what we found previously for another non-thermal SNR RX J1713.7-3946. 
\item Even if the inconsistency with the thermal X-ray line emission is ignored, the hadronic model, with the SNR forward shock moving into a uniform ISM, fails to reproduce the spectral index of the observed X-ray synchrotron emission, as measured by  \textit{ASCA} GIS and 
\textit{Chandra} ACIS.  While the hadronic model can produce a reasonable match to the spectral shapes of data in other wavelengths, the  overall normalization of the broadband emission requires a boosting factor of $\sim 5$ to bring the model spectra to the observed flux levels.
\item Another possible problem with the  hadronic model is that our best fit hadronic result predicts a spatial (i.e., radial) 
variation of the X-ray synchrotron photon index around the FS that is incompatible to the trend suggested by data from X-ray satellites
(see Figure~\ref{fig:synch_index}).
%
%
\item The Vela Jr.~SNR most likely originated from a core-collapse SN but the  ``best-fit'' hadronic model requires a relatively high-density, 
high-magnetic field environment more typical of a thermonuclear Type Ia SN.
\end{itemize}

In contrast to the above problems with the hadronic model, our pre-SN wind leptonic model can \SCly\ explain the broadband continuum emission, the lack of thermal X-ray line emission, the overall normalization assuming $\dSNR \sim 1$\,kpc, and the spatial variation of the non-thermal 
X-ray spectral index. 

It is important to note that while our calculations assume a uniform ambient ISM,  any real SNR may have a highly inhomogeneous upstream medium containing clumped gas.
Recently, radio observations of SNR RX J1713.7-3946 \citep{Sano2010, Fukui2012} have revealed clumpy atomic/molecular cloud distributions surrounding the SNR shell, indicating a possible interaction of the shock and the accelerated protons with the gas clouds. Concentrating the mass in small, dense clumps can considerably enhance the gamma-ray flux. 
\citet{Inoue2012} have investigated this possibility using 3-D MHD simulations of a shock running into an inhomogeneous medium.
They  pointed out the possibility that a hadronic model can avoid the problem of an over-prediction of the thermal X-ray flux, since the denser cores of the gas that carry most of the mass fraction can survive the shock passing without being ionized and hence will not contribute to the thermal emission, but still can interact with the accelerated protons and produce gamma-rays. 
While definitively ruling out hadronic scenarios in homogeneous broadband models of Vela Jr., our 1-D model cannot rule out hadronic dominated GeV-TeV emission in more complex clumpy environments.
In fact, although not pursued in this work, it is possible that the gamma-rays are contributed by a two-component mixture of IC and \pdecay\ photons from the interaction of dense gas clumps and the accelerated CR protons. Future high-resolution observations of the surrounding interstellar gas distribution and 3-D hydro simulations self-consistently coupled to DSA will provide a quantitative estimate on this possible \pdecay\ component from clumps. 

To explain the broadband spectrum of the SNR as a whole, our leptonic model requires a low B-field ($\sim$ a few $\mu$G)  on average in the SNR shell, which is one of the reasons why the presence of a wind cavity is necessary to provide a weak B-field in the upstream medium. The fact that magnetic field amplification in the shock precursor is taken into account in our models only strengthens this need. This low downstream B-field is highly consistent with those inferred from measurements of the X-ray/TeV brightness ratio \citep{Aharonian2007b} and more recently the radial variation of X-ray synchrotron spectral indices \citep{Pannuti2010, KHU2013}. However, as briefly mentioned earlier, it disagrees with some previous claims of large post-shock B-fields ($\sim 0.1$~mG) based on the discovery of sharp X-ray filaments in the NW shell. The argument for such large field strength is built on the assumption that the narrow widths of such filaments are realized by a short synchrotron loss time-scale for the electrons. Our broadband analysis and supporting observational data essentially show that an alternative explanation for the sharp filamentary structures without the need to invoke large B-field strengths may be necessary \citep[see e.g.,][and references therein]{BykovDots2008,EV2008}. We leave this important point as an open question to be answered by future works. 
%

A fundamental question for DSA in SNRs is the fraction of SN explosion energy, $\fSN$, channeled into CRs. Our two models show that this fraction can be very different, {\it at a given age}, depending on whether the gamma-ray production mechanism is assumed to be \pdecay\ or IC. For Vela jr, at $\tSNR=2500$\,yr, $\fSN \simeq 0.14$ for our leptonic model and $\fSN \simeq 0.48$ for our hadronic model. In both cases, of course, CR protons contain the large majority of the CR energy. The ratio of CR electron to CR proton energy density is given by $\Kep$ and $\Kep \simeq 0.015$ for our leptonic model and $\Kep \simeq 1.5\xx{-4}$ for our hadronic model.

The determination of the gamma-ray origin, therefore, relates directly to the long postulated but unproved link between SNRs and Galactic CRs, that is, whether or not the Galactic ensemble of SNRs can meet the energy budget of the CR spectrum. Because the \pdecay\ and IC mechanisms are so different, a mis-interpretation of the 
multi-wavelength data including the gamma-rays can yield a very wrong prediction for the contribution of SNRs to the Galactic CRs.
Of course, the CR energy budget depends ultimately on $\fSN$ over the full age of the remnant so it will be interesting to see how $\fSN$ evolves from very young ages up to the radiative phase of a SNR.
In particular, the recent discovery of gamma-ray bright middle-aged SNRs by Fermi LAT may imply active particle acceleration at radiative shocks running into molecular clouds despite their slow speeds.

Another important point is that the evolution of a SNR in a changing ambient environment may shift the dominating gamma-ray production mechanism with age. It is hence critical to understand observations from SNRs of different ages and try to link them together into a self-consistent evolutionary picture. We will pursue this line of work in a series of follow-up papers. 

Finally, we discussed the implications of our models for Vela Jr.~on future observations by the next-generation of observatories, including \textit{CTA} and \textit{Astro-H}.
We showed that \textit{CTA} may be able to distinguish broadband emission models by measuring TeV brightness profiles with its unprecedented angular resolving power. This is especially valuable to distinguish leptonic and hadronic models for Vela Jr., as we have illustrated in 
Figure~\ref{fig:TeV_profile}. 
In the X-ray band, the soon to be commissioned \textit{Astro-H} space telescope possesses an extremely powerful calorimeter for 
high-resolution spectral studies, and we have postulated that there is a high possibility that SXS will be able to detect the more prominent thermal 
lines (see Figure~\ref{fig:spec_astro_h}) from non-thermal dominated SNRs such as Vela Jr.~and RX J1713.7-3946. We expect future multi-wavelength observations will put extremely precise constraints on our model parameters, allowing us to firmly pin down the origin of the high-energy emission and to gain further insight on the particle acceleration mechanism in young SNR shocks.   

\acknowledgments 
The authors acknowledge important discussions with Takaaki Tanaka and Hidetoshi Sano concerning this work. 
They also thank the anonymous referee for providing valuable suggestions on improving the quality of this manuscript. 
D.C.E. acknowledges support from NASA grants ATP02-0042-0006, NNH04Zss001N-LTSA, 06-ATP06-21, and NNX11AE03G. 
P.S. acknowledges support from NASA Contract NAS8-03060. 
S.N. acknowledges support from Ministry of Education, Culture, Sports, Science and Technology (No. 23105709), Japan Society for the Promotion of Science (No. 19104006 and No. 23340069), and the Global COE Program, The Next Generation of Physics, Spun from Universality and Emergence, from MEXT of Japan.  

\bibliographystyle{aa} 
\bibliography{reference}

\begin{thebibliography}{38}
\expandafter\ifx\csname natexlab\endcsname\relax\def\natexlab#1{#1}\fi

\bibitem[{{Abdo} {et~al.}(2009){Abdo}, {Ackermann}, {Ajello}, {Baldini},
  {Ballet}, {Barbiellini}, {Baring}, {Bastieri}, {Baughman}, {Bechtol},
  {Bellazzini}, {Berenji}, {Blandford}, {Bloom}, {Bonamente}, {Borgland},
  {Bouvier}, {Bregeon}, {Brez}, {Brigida}, {Bruel}, {Burnett}, {Buson},
  {Caliandro}, {Cameron}, {Caraveo}, {Casandjian}, {Cecchi}, {{\c C}elik},
  {Chekhtman}, {Cheung}, {Chiang}, {Ciprini}, {Claus}, {Cohen-Tanugi},
  {Cominsky}, {Conrad}, {Cutini}, {Dermer}, {de Angelis}, {de Palma}, {Digel},
  {Dormody}, {Silva}, {Drell}, {Dubois}, {Dumora}, {Farnier}, {Favuzzi},
  {Fegan}, {Focke}, {Fortin}, {Frailis}, {Fukazawa}, {Funk}, {Fusco},
  {Gargano}, {Gasparrini}, {Gehrels}, {Germani}, {Giavitto}, {Giebels},
  {Giglietto}, {Giordano}, {Glanzman}, {Godfrey}, {Grenier}, {Grondin},
  {Grove}, {Guillemot}, {Guiriec}, {Hanabata}, {Harding}, {Hayashida}, {Hays},
  {Hughes}, {Jackson}, {J{\'o}hannesson}, {Johnson}, {Johnson}, {Johnson},
  {Kamae}, {Katagiri}, {Kataoka}, {Katsuta}, {Kawai}, {Kerr}, {Kn{\"o}dlseder},
  {Kocian}, {Kuss}, {Lande}, {Latronico}, {Lemoine-Goumard}, {Longo},
  {Loparco}, {Lott}, {Lovellette}, {Lubrano}, {Makeev}, {Mazziotta}, {McEnery},
  {Meurer}, {Michelson}, {Mitthumsiri}, {Mizuno}, {Moiseev}, {Monte},
  {Monzani}, {Morselli}, {Moskalenko}, {Murgia}, {Nakamori}, {Nolan}, {Norris},
  {Nuss}, {Ohsugi}, {Okumura}, {Omodei}, {Orlando}, {Ormes}, {Paneque},
  {Parent}, {Pelassa}, {Pepe}, {Pesce-Rollins}, {Piron}, {Porter}, {Rain{\`o}},
  {Rando}, {Razzano}, {Reimer}, {Reimer}, {Reposeur}, {Ritz}, {Rodriguez},
  {Romani}, {Roth}, {Ryde}, {Sadrozinski}, {Sanchez}, {Sander}, {Saz
  Parkinson}, {Scargle}, {Schalk}, {Sgr{\`o}}, {Siskind}, {Smith}, {Smith},
  {Spandre}, {Spinelli}, {Strickman}, {Suson}, {Tajima}, {Takahashi},
  {Takahashi}, {Tanaka}, {Thayer}, {Thayer}, {Thompson}, {Tibaldo}, {Tibolla},
  {Torres}, {Tosti}, {Tramacere}, {Uchiyama}, {Usher}, {Vasileiou}, {Venter},
  {Vilchez}, {Vitale}, {Waite}, {Wang}, {Winer}, {Wood}, {Yamazaki}, {Ylinen},
  \& {Ziegler}}]{AbdoEtalW51C2009}
{Abdo}, A.~A., {Ackermann}, M., {Ajello}, M., {et~al.} 2009, \apjl, 706, L1

\bibitem[{{Abdo} {et~al.}(2010{\natexlab{a}}){Abdo}, {Ackermann}, {Ajello},
  {Baldini}, {Ballet}, {Barbiellini}, {Baring}, {Bastieri}, {Baughman},
  {Bechtol}, {Bellazzini}, {Berenji}, {Blandford}, {Bloom}, {Bonamente},
  {Borgland}, {Bregeon}, {Brez}, {Brigida}, {Bruel}, {Burnett}, {Buson},
  {Caliandro}, {Cameron}, {Caraveo}, {Casandjian}, {Cecchi}, {{\c C}elik},
  {Chekhtman}, {Cheung}, {Chiang}, {Ciprini}, {Claus}, {Cognard},
  {Cohen-Tanugi}, {Cominsky}, {Conrad}, {Cutini}, {Dermer}, {de Angelis}, {de
  Palma}, {Digel}, {do Couto e Silva}, {Drell}, {Dubois}, {Dumora}, {Espinoza},
  {Farnier}, {Favuzzi}, {Fegan}, {Focke}, {Fortin}, {Frailis}, {Fukazawa},
  {Funk}, {Fusco}, {Gargano}, {Gasparrini}, {Gehrels}, {Germani}, {Giavitto},
  {Giebels}, {Giglietto}, {Giordano}, {Glanzman}, {Godfrey}, {Grenier},
  {Grondin}, {Grove}, {Guillemot}, {Guiriec}, {Hanabata}, {Harding},
  {Hayashida}, {Hays}, {Hughes}, {Jackson}, {J{\'o}hannesson}, {Johnson},
  {Johnson}, {Johnson}, {Kamae}, {Katagiri}, {Kataoka}, {Katsuta}, {Kawai},
  {Kerr}, {Kn{\"o}dlseder}, {Kocian}, {Kramer}, {Kuss}, {Lande}, {Latronico},
  {Lemoine-Goumard}, {Longo}, {Loparco}, {Lott}, {Lovellette}, {Lubrano},
  {Lyne}, {Madejski}, {Makeev}, {Mazziotta}, {McEnery}, {Meurer}, {Michelson},
  {Mitthumsiri}, {Mizuno}, {Monte}, {Monzani}, {Morselli}, {Moskalenko},
  {Murgia}, {Nakamori}, {Nolan}, {Norris}, {Noutsos}, {Nuss}, {Ohsugi},
  {Omodei}, {Orlando}, {Ormes}, {Paneque}, {Parent}, {Pelassa}, {Pepe},
  {Pesce-Rollins}, {Piron}, {Porter}, {Rain{\`o}}, {Rando}, {Razzano},
  {Reimer}, {Reimer}, {Reposeur}, {Rochester}, {Rodriguez}, {Romani}, {Roth},
  {Ryde}, {Sadrozinski}, {Sanchez}, {Sander}, {Parkinson}, {Scargle},
  {Sgr{\`o}}, {Siskind}, {Smith}, {Smith}, {Spandre}, {Spinelli}, {Stappers},
  {Stecker}, {Strickman}, {Suson}, {Tajima}, {Takahashi}, {Takahashi},
  {Tanaka}, {Thayer}, {Thayer}, {Theureau}, {Thompson}, {Tibaldo}, {Tibolla},
  {Torres}, {Tosti}, {Tramacere}, {Uchiyama}, {Usher}, {Vasileiou}, {Venter},
  {Vilchez}, {Vitale}, {Waite}, {Wang}, {Winer}, {Wood}, {Yamazaki}, {Ylinen},
  \& {Ziegler}}]{AbdoEtalW442010}
{Abdo}, A.~A., {Ackermann}, M., {Ajello}, M., {et~al.} 2010{\natexlab{a}},
  Science, 327, 1103

\bibitem[{{Abdo} {et~al.}(2010{\natexlab{b}}){Abdo}, {Ackermann}, {Ajello},
  {Baldini}, {Ballet}, {Barbiellini}, {Bastieri}, {Baughman}, {Bechtol},
  {Bellazzini}, {Berenji}, {Blandford}, {Bloom}, {Bonamente}, {Borgland},
  {Bregeon}, {Brez}, {Brigida}, {Bruel}, {Burnett}, {Buson}, {Caliandro},
  {Cameron}, {Caraveo}, {Casandjian}, {Cecchi}, {{\c C}elik}, {Chekhtman},
  {Cheung}, {Chiang}, {Cillis}, {Ciprini}, {Claus}, {Cohen-Tanugi}, {Cominsky},
  {Conrad}, {Cutini}, {Dermer}, {de Angelis}, {de Palma}, {Silva}, {Drell},
  {Drlica-Wagner}, {Dubois}, {Dumora}, {Farnier}, {Favuzzi}, {Fegan}, {Focke},
  {Fortin}, {Frailis}, {Fukazawa}, {Funk}, {Fusco}, {Gargano}, {Gasparrini},
  {Gehrels}, {Germani}, {Giavitto}, {Giebels}, {Giglietto}, {Giordano},
  {Glanzman}, {Godfrey}, {Grenier}, {Grondin}, {Grove}, {Guillemot}, {Guiriec},
  {Hanabata}, {Harding}, {Hayashida}, {Hughes}, {Jackson}, {J{\'o}hannesson},
  {Johnson}, {Johnson}, {Johnson}, {Kamae}, {Katagiri}, {Kataoka}, {Kawai},
  {Kerr}, {Kn{\"o}dlseder}, {Kocian}, {Kuss}, {Lande}, {Latronico}, {Lee},
  {Lemoine-Goumard}, {Longo}, {Loparco}, {Lott}, {Lovellette}, {Lubrano},
  {Madejski}, {Makeev}, {Mazziotta}, {Meurer}, {Michelson}, {Mitthumsiri},
  {Moiseev}, {Monte}, {Monzani}, {Morselli}, {Moskalenko}, {Murgia},
  {Nakamori}, {Nolan}, {Norris}, {Nuss}, {Ohsugi}, {Orlando}, {Ormes}, {Ozaki},
  {Paneque}, {Panetta}, {Parent}, {Pelassa}, {Pepe}, {Pesce-Rollins}, {Piron},
  {Porter}, {Rain{\`o}}, {Rando}, {Razzano}, {Reimer}, {Reimer}, {Reposeur},
  {Rochester}, {Rodriguez}, {Romani}, {Roth}, {Ryde}, {Sadrozinski}, {Sanchez},
  {Sander}, {Saz Parkinson}, {Scargle}, {Sgr{\`o}}, {Siskind}, {Smith},
  {Smith}, {Spandre}, {Spinelli}, {Strickman}, {Strong}, {Suson}, {Tajima},
  {Takahashi}, {Takahashi}, {Tanaka}, {Thayer}, {Thayer}, {Thompson},
  {Tibaldo}, {Torres}, {Tosti}, {Tramacere}, {Uchiyama}, {Usher}, {Van Etten},
  {Vasileiou}, {Venter}, {Vilchez}, {Vitale}, {Waite}, {Wang}, {Winer}, {Wood},
  {Ylinen}, \& {Ziegler}}]{AbdoEtalIC4432010}
{Abdo}, A.~A., {Ackermann}, M., {Ajello}, M., {et~al.} 2010{\natexlab{b}},
  \apj, 712, 459

\bibitem[{{Aharonian} {et~al.}(2007){Aharonian}, {Akhperjanian}, {Bazer-Bachi},
  {Beilicke}, {Benbow}, {Berge}, {Bernl{\"o}hr}, {Boisson}, {Bolz}, {Borrel},
  {Braun}, {Brown}, {B{\"u}hler}, {B{\"u}sching}, {Carrigan}, {Chadwick},
  {Chounet}, {Coignet}, {Cornils}, {Costamante}, {Degrange}, {Dickinson},
  {Djannati-Ata{\"i}}, {Drury}, {Dubus}, {Egberts}, {Emmanoulopoulos},
  {Espigat}, {Feinstein}, {Ferrero}, {Fiasson}, {Filipovic}, {Fontaine},
  {Fukui}, {Funk}, {Funk}, {F{\"u}{\ss}ling}, {Gallant}, {Giebels},
  {Glicenstein}, {Goret}, {Hadjichristidis}, {Hauser}, {Hauser}, {Heinzelmann},
  {Henri}, {Hermann}, {Hinton}, {Hiraga}, {Hoffmann}, {Hofmann}, {Holleran},
  {Hoppe}, {Horns}, {Ishisaki}, {Jacholkowska}, {de Jager}, {Kendziorra},
  {Kerschhaggl}, {Kh{\'e}lifi}, {Komin}, {Konopelko}, {Kosack}, {Lamanna},
  {Latham}, {Le Gallou}, {Lemi{\`e}re}, {Lemoine-Goumard}, {Lohse}, {Martin},
  {Martineau-Huynh}, {Marcowith}, {Masterson}, {Maurin}, {McComb}, {Moulin},
  {Moriguchi}, {de Naurois}, {Nedbal}, {Nolan}, {Noutsos}, {Orford}, {Osborne},
  {Ouchrif}, {Panter}, {Pelletier}, {Pita}, {P{\"u}hlhofer}, {Punch},
  {Ranchon}, {Raubenheimer}, {Raue}, {Rayner}, {Reimer}, {Ripken}, {Rob},
  {Rolland}, {Rosier-Lees}, {Rowell}, {Sahakian}, {Santangelo}, {Saug{\'e}},
  {Schlenker}, {Schlickeiser}, {Schr{\"o}der}, {Schwanke}, {Schwarzburg},
  {Schwemmer}, {Shalchi}, {Sol}, {Spangler}, {Spanier}, {Steenkamp},
  {Stegmann}, {Superina}, {Tam}, {Tavernet}, {Terrier}, {Tluczykont}, {van
  Eldik}, {Vasileiadis}, {Venter}, {Vialle}, {Vincent}, {V{\"o}lk}, {Wagner},
  \& {Ward}}]{Aharonian2007b}
{Aharonian}, F., {Akhperjanian}, A.~G., {Bazer-Bachi}, A.~R., {et~al.} 2007,
  \apj, 661, 236

\bibitem[{{Aschenbach}(1998)}]{Aschenbach1998}
{Aschenbach}, B. 1998, \nat, 396, 141

\bibitem[{{Bamba} {et~al.}(2005){Bamba}, {Yamazaki}, \& {Hiraga}}]{Bamba2005}
{Bamba}, A., {Yamazaki}, R., \& {Hiraga}, J.~S. 2005, \apj, 632, 294

\bibitem[{{Bell}(2004)}]{Bell2004}
{Bell}, A.~R. 2004, \mnras, 353, 550

\bibitem[{{Berezhko} {et~al.}(2009){Berezhko}, {P{\"u}hlhofer}, \&
  {V{\"o}lk}}]{Berezhko2009}
{Berezhko}, E.~G., {P{\"u}hlhofer}, G., \& {V{\"o}lk}, H.~J. 2009, \aap, 505,
  641

\bibitem[{{Blasi} {et~al.}(2005){Blasi}, {Gabici}, \& {Vannoni}}]{BGV2005}
{Blasi}, P., {Gabici}, S., \& {Vannoni}, G. 2005, \mnras, 361, 907

\bibitem[{{Blondin} \& {Ellison}(2001)}]{BE2001}
{Blondin}, J.~M. \& {Ellison}, D.~C. 2001, \apj, 560, 244

\bibitem[{{Bykov} {et~al.}(2011){Bykov}, {Osipov}, \& {Ellison}}]{BOE2011}
{Bykov}, A.~M., {Osipov}, S.~M., \& {Ellison}, D.~C. 2011, \mnras, 410, 39

\bibitem[{{Bykov} {et~al.}(2008){Bykov}, {Uvarov}, \&
  {Ellison}}]{BykovDots2008}
{Bykov}, A.~M., {Uvarov}, Y.~A., \& {Ellison}, D.~C. 2008, \apjl, 689, L133

\bibitem[{{Caprioli} {et~al.}(2009){Caprioli}, {Blasi}, {Amato}, \&
  {Vietri}}]{CBAV2009}
{Caprioli}, D., {Blasi}, P., {Amato}, E., \& {Vietri}, M. 2009, \mnras, 395,
  895

\bibitem[{{Combi} {et~al.}(1999){Combi}, {Romero}, \& {Benaglia}}]{Combi1999}
{Combi}, J.~A., {Romero}, G.~E., \& {Benaglia}, P. 1999, \apjl, 519, L177

\bibitem[{{Ellison}(2000)}]{Ellison2000}
{Ellison}, D.~C. 2000, AIP Conf. Proc. 528, Acceleration and Transport of
  Energetic Particles Observed in the Heliosphere, ed. R. A. Mewaldt, J. R.
  Jokipii, M. A. Lee, E. M\"obius, \& T. H. Zurbuchen (New York: AIP), 386

\bibitem[{{Ellison} {et~al.}(2007){Ellison}, {Patnaude}, {Slane}, {Blasi}, \&
  {Gabici}}]{EPSBG2007}
{Ellison}, D.~C., {Patnaude}, D.~J., {Slane}, P., {Blasi}, P., \& {Gabici}, S.
  2007, \apj, 661, 879

\bibitem[{{Ellison} {et~al.}(2010){Ellison}, {Patnaude}, {Slane}, \&
  {Raymond}}]{EPSR2010}
{Ellison}, D.~C., {Patnaude}, D.~J., {Slane}, P., \& {Raymond}, J. 2010, \apj,
  712, 287

\bibitem[{{Ellison} {et~al.}(2012){Ellison}, {Slane}, {Patnaude}, \&
  {Bykov}}]{ESPB2012}
{Ellison}, D.~C., {Slane}, P., {Patnaude}, D.~J., \& {Bykov}, A.~M. 2012, \apj,
  744, 39

\bibitem[{{Ellison} \& {Vladimirov}(2008)}]{EV2008}
{Ellison}, D.~C. \& {Vladimirov}, A. 2008, \apjl, 673, L47

\bibitem[{{Fukui} {et~al.}(2012){Fukui}, {Sano}, {Sato}, {Torii}, {Horachi},
  {Hayakawa}, {McClure-Griffiths}, {Rowell}, {Inoue}, {Inutsuka}, {Kawamura},
  {Yamamoto}, {Okuda}, {Mizuno}, {Onishi}, {Mizuno}, \& {Ogawa}}]{Fukui2012}
{Fukui}, Y., {Sano}, H., {Sato}, J., {et~al.} 2012, \apj, 746, 82

\bibitem[{{Hiraga} {et~al.}(2009){Hiraga}, {Kobayashi}, {Tamagawa}, {Hayato},
  {Bamba}, {Terada}, {Petre}, {Katagiri}, \& {Tsunemi}}]{Hiraga2009}
{Hiraga}, J.~S., {Kobayashi}, Y., {Tamagawa}, T., {et~al.} 2009, \pasj, 61, 275

\bibitem[{{Inoue} {et~al.}(2012){Inoue}, {Yamazaki}, {Inutsuka}, \&
  {Fukui}}]{Inoue2012}
{Inoue}, T., {Yamazaki}, R., {Inutsuka}, S.-i., \& {Fukui}, Y. 2012, \apj, 744,
  71

\bibitem[{{Iyudin} {et~al.}(1998){Iyudin}, {Sch{\"o}nfelder}, {Bennett},
  {Bloemen}, {Diehl}, {Hermsen}, {Lichti}, {van der Meulen}, {Ryan}, \&
  {Winkler}}]{Iyudin1998}
{Iyudin}, A.~F., {Sch{\"o}nfelder}, V., {Bennett}, K., {et~al.} 1998, \nat,
  396, 142

\bibitem[{{Kamae} {et~al.}(2006){Kamae}, {Karlsson}, {Mizuno}, {Abe}, \&
  {Koi}}]{Kamae06}
{Kamae}, T., {Karlsson}, N., {Mizuno}, T., {Abe}, T., \& {Koi}, T. 2006, \apj,
  647, 692

\bibitem[{{Kargaltsev} {et~al.}(2002){Kargaltsev}, {Pavlov}, {Sanwal}, \&
  {Garmire}}]{Kargaltsev2002}
{Kargaltsev}, O., {Pavlov}, G.~G., {Sanwal}, D., \& {Garmire}, G.~P. 2002,
  \apj, 580, 1060

\bibitem[{{Katsuda} {et~al.}(2008){Katsuda}, {Tsunemi}, \&
  {Mori}}]{Katsuda2008}
{Katsuda}, S., {Tsunemi}, H., \& {Mori}, K. 2008, \apjl, 678, L35

\bibitem[{{Kishishita} {et~al.}(2013){Kishishita}, {Hiraga}, \&
  {Uchiyama}}]{KHU2013}
{Kishishita}, T., {Hiraga}, J., \& {Uchiyama}, Y. 2013, \aap, in press

\bibitem[{{Lee} {et~al.}(2012){Lee}, {Ellison}, \& {Nagataki}}]{LEN2012}
{Lee}, S.-H., {Ellison}, D.~C., \& {Nagataki}, S. 2012, \apj, 750, 156

\bibitem[{{Pannuti} {et~al.}(2010){Pannuti}, {Allen}, {Filipovi{\'c}}, {De
  Horta}, {Stupar}, \& {Agrawal}}]{Pannuti2010}
{Pannuti}, T.~G., {Allen}, G.~E., {Filipovi{\'c}}, M.~D., {et~al.} 2010, \apj,
  721, 1492

\bibitem[{{Patnaude} {et~al.}(2009){Patnaude}, {Ellison}, \& {Slane}}]{PES2009}
{Patnaude}, D.~J., {Ellison}, D.~C., \& {Slane}, P. 2009, \apj, 696, 1956

\bibitem[{{Patnaude} {et~al.}(2010){Patnaude}, {Slane}, {Raymond}, \&
  {Ellison}}]{PSRE2010}
{Patnaude}, D.~J., {Slane}, P., {Raymond}, J.~C., \& {Ellison}, D.~C. 2010,
  \apj, 725, 1476

\bibitem[{{Porter} {et~al.}(2006){Porter}, {Moskalenko}, \& {Strong}}]{PMS2006}
{Porter}, T.~A., {Moskalenko}, I.~V., \& {Strong}, A.~W. 2006, \apjl, 648, L29

\bibitem[{{Sano} {et~al.}(2010){Sano}, {Sato}, {Horachi}, {Moribe}, {Yamamoto},
  {Hayakawa}, {Torii}, {Kawamura}, {Okuda}, {Mizuno}, {Onishi}, {Maezawa},
  {Inoue}, {Inutsuka}, {Tanaka}, {Matsumoto}, {Mizuno}, {Ogawa}, {Stutzki},
  {Bertoldi}, {Anderl}, {Bronfman}, {Koo}, {Burton}, {Benz}, \&
  {Fukui}}]{Sano2010}
{Sano}, H., {Sato}, J., {Horachi}, H., {et~al.} 2010, \apj, 724, 59

\bibitem[{{Sch{\"o}nfelder} {et~al.}(2000){Sch{\"o}nfelder}, {Bloemen},
  {Collmar}, {Diehl}, {Hermsen}, {Kn{\"o}dlseder}, {Lichti}, {Pl{\"u}schke},
  {Ryan}, {Strong}, \& {Winkler}}]{Schonfelder2000}
{Sch{\"o}nfelder}, V., {Bloemen}, H., {Collmar}, W., {et~al.} 2000, in American
  Institute of Physics Conference Series, Vol. 510, American Institute of
  Physics Conference Series, ed. M.~L. {McConnell} \& J.~M. {Ryan}, 54--59

\bibitem[{{Slane} {et~al.}(2001){Slane}, {Hughes}, {Edgar}, {Plucinsky},
  {Miyata}, {Tsunemi}, \& {Aschenbach}}]{Slane2001}
{Slane}, P., {Hughes}, J.~P., {Edgar}, R.~J., {et~al.} 2001, \apj, 548, 814

\bibitem[{{Stupar} {et~al.}(2005){Stupar}, {Filipovi{\'c}}, {Jones}, \&
  {Parker}}]{Stupar2005}
{Stupar}, M., {Filipovi{\'c}}, M.~D., {Jones}, P.~A., \& {Parker}, Q.~A. 2005,
  Advances in Space Research, 35, 1047

\bibitem[{{Tanaka} {et~al.}(2011){Tanaka}, {Allafort}, {Ballet}, {Funk},
  {Giordano}, {Hewitt}, {Lemoine-Goumard}, {Tajima}, {Tibolla}, \&
  {Uchiyama}}]{Tanaka2011}
{Tanaka}, T., {Allafort}, A., {Ballet}, J., {et~al.} 2011, \apjl, 740, L51

\bibitem[{{Vladimirov} {et~al.}(2008){Vladimirov}, {Bykov}, \&
  {Ellison}}]{VBE2008}
{Vladimirov}, A.~E., {Bykov}, A.~M., \& {Ellison}, D.~C. 2008, \apj, 688, 1084

\end{thebibliography}

\newpage

\begin{center} 
\begin{deluxetable*}{llll} 
\tablecolumns{4}
\tablewidth{14cm}
\tablecaption{Model Summary \tnac}
\tablehead{\colhead{} & \colhead{Hadronic \tna} & \colhead{Leptonic \tnb}  &\colhead{Remarks}}
\startdata
\sidehead{Input parameters}
$d_\mrm{SNR}$ [kpc]		& 0.74	& 0.88		& Distance to SNR\\
$n_0$ [cm$^{-3}$]	& 0.033	 	& (0.002)	& Far upstream gas density\\
$B_0$ [$\mu$G]	& 0.5 	 	& (0.14)	& Far upstream B-field\\
$\dMdt$ [$10^{-6}$ \SunMyr]	& $-$	& $7.5$	& Pre-SN mass loss rate \\
$\Vwind$ [\kmps]	& $-$	& 50 & Pre-SN wind speed \\
$\SigWind$		& $-$	& 0.02	& Wind magnetization \\
$\Kep$ 	& $1.5 \times 10^{-4}$ 	& $0.015$  & (e$^\mrm{-}$/p) ratio at $\pmax$ of e$^\mrm{-}$\\
$\cutoff$	& 0.75	& 0.50		& Exp. cut-off index \tnc\\
$f_\mrm{FEB}$	& $0.15$	& $0.12$	& Width of FEB over $R_\mrm{FS}$ \tnd\\
$\falf$	& $0.10$	& $1.00$	& $x$-dependance of $\Valf$ \tne\\
\\
\hline
\sidehead{Output quantities}
$R_\mrm{FS}$	[pc]	& 12.7	& 15.2	& Forward shock radius \\
$R_\mrm{CD}$	[pc]	& 10.3	& 12.5	& Contact discontinuity radius \\
$V_\mrm{FS}$ [\kmps]	& 2130	& 4700	& Forward shock velocity\\
$\pmax$ (p) [TeV/c] 	& 26.7	& 5.2 	& Proton max. momentum\\
$\pmax$ (e$^\mrm{-}$) [TeV/c] 	& 13.3	& 5.2		& Electron max. momentum\\
$R_\mrm{tot}$		& 9.30	& 4.69 	& Total compression ratio\\
$R_\mrm{sub}$	& 3.69	& 3.99	& Subshock compression ratio\\
$B_2$ [$\mu$G]	& 34.1	& 4.8		& Downstream B-field\\
$T_2$ [$10^8$ K]	& 0.16	& 3.62	& Downstream temperature\\
$\epsilon_{acc}$	& 0.84	& 0.36	& Total accel efficiency\\
$\epsilon_{esc}$	& 0.34	& 0.12	& Escaped accel efficiency\\
$E_\mrm{CR}/\EnSN$ ($f_\mrm{SN}$)	& 0.48	& 0.14	& SN energy converted to CR
\enddata
\tablenotetext{$\ast$}{For all models listed in this Table, Bohm diffusion is assumed in the DS and precursor region, and MFA is included.
These models all have $\EnSN=10^{51}$\,erg,  $T_0 = 10^4$\,K, and $\fHe=0.0977$.}
\tablenotetext{$\dagger $}{Values in parentheses are taken at $t = t_\mrm{age}$ (= 2500~yr).}
\tablenotetext{a}{Best-fit model under the `hadronic-dominated' scenario with a uniform upstream medium.} 
\tablenotetext{b}{Best-fit model under the `leptonic-dominated' scenario with a pre-SN magnetized wind.} 
\tablenotetext{c}{See e.g. equation~(24) in \citet{LEN2012} for definition.}
\tablenotetext{d}{See e.g. equation~(1) and description in \citet{EV2008}.}
\tablenotetext{e}{See e.g. equation~(14) in \citet{LEN2012} for definition.}

\label{table:param}  
\end{deluxetable*}
\end{center}

\end{document}